\documentclass[aps,amssymb,amsmath,superscriptaddress,twocolumn,showpacs,floatfix]{revtex4-1}
\usepackage{graphicx}
\usepackage{subfigure}
\usepackage{amsfonts}
\usepackage{extarrows}
\usepackage{geometry}
\usepackage{ulem}

\begin{document}
\title{Phase diagram of the spin-1/2 Heisenberg alternating chain in a magnetic field}
\author{Wei-Xia Chen}
\affiliation{School of Physical Science and Technology, Soochow University, Suzhou, Jiangsu 215006, China}
\affiliation{Department of Physics and Jiangsu Laboratory of Advanced
Functional Material, Changshu Institute of Technology, Changshu 215500, China}
\author{Jie Ren}
\email{jren@cslg.edu.cn}
\affiliation{Department of Physics and Jiangsu Laboratory of Advanced
Functional Material, Changshu Institute of Technology, Changshu 215500, China}
\author{Wen-Long You}
\email{wlyou@suda.edu.cn}
\affiliation{School of Physical Science and Technology, Soochow University, Suzhou, Jiangsu 215006, China}

\author{Xiang Hao}
\affiliation{Department of Physics, School of Mathematics and Physics, Suzhou University of
Science and Technology, Suzhou, Jiangsu 215009, Peoples Republic of China.}
\author{Yin-Zhong Wu}
\affiliation{Department of Physics, School of Mathematics and Physics, Suzhou University of
Science and Technology, Suzhou, Jiangsu 215009, Peoples Republic of China.}

\date{\today}
\begin{abstract}
By using the infinite time-evolving block decimation, we study quantum fidelity and entanglement entropy in the spin-1/2 Heisenberg alternating chain under an external magnetic field. The effects of the magnetic field on the fidelity are investigated, and its relation with the quantum phase transition (QPT) is analyzed. The phase diagram of the model is given accordingly, which supports the Haldane phase, the singlet-dimer phase, the Luttinger liquid phase and the paramagnetic phase. The scaling of entanglement entropy in the gapless Luttinger liquid phase is studied, and the central charge $c=1$ is obtained. We also study the relationship between the quantum coherence, string order parameter and QPTs. Results obtained from these quantum information observations are consistent with the previous reports.
\end{abstract}
\maketitle

\section{introduction}
\label{sec:intorduction}
Quantum phase transition (QPT) is a purely quantum
process occurring in strongly correlated many-body systems at absolute zero temperature due to quantum fluctuations~\cite{Sachdev}.
The spin chains attract a lot of attention since they give rise to many exotic properties in the ground state, such as bond alternating spin-1/2 Heisenberg chain~\cite{Yu,FHM,Orignac,Klyushina,James,Tennant,xiong,HTW}.  
A number of compounds are discovered whose properties can be explained by invoking bond alternating chains.
LiInCr$_4$O$_8$ was found to be spin-3/2 breathing pyrochlore antiferromagnet, which is an alternating array of small and large tetrahedra \cite{Okamoto17}. Recently, it is reported that the results from the dimer
anisotropic XYZ model are relevant to a large number of quasi-one dimensional magnets \cite{Xu}.  In history, Bulaevskii predicted that a spin gap exists in the nonuniform antiferromagnetic (AFM) spin chains~\cite{Bulaevskii}.
Kohmoto found the existence of the Haldane phase synonymous with hidden $D_2$ symmetry breaking in the AFM-ferromagnetic~(FM) bond alternating spin-1/2 Heisenberg chain~\cite{Kohmoto}. Furthermore, it was  pointed out that isotropic $S=1/2$ Heisenberg chain with alternating AFM and FM couplings can be mapped onto the isotropic $S=1$ AFM Heisenberg chain when the FM couplings tend to infinity~\cite{Hida}. The compounds like CuGeO$_3$~\cite{Hase} exhibiting spin-Peierls transitions belong to AFM-AFM bond
alternating class, and DMACuCl$_3$~\cite{Stone} was claimed to fall into the $S=1/2$ AFM-FM bond alternating class.
Additionally, the quantum simulation using ultracold atoms systems \cite{Bloch12,Sun14,Sun2014} and  trapped
polariton condensates \cite{Ohadi17} has made great progress
in creating interesting quantum models motivated by solid-state physics. Geometrically frustrated magnets such as zigzag chains
can be designed and tuned by the depth of the optical lattice, and thus
nonuniform configurations in the ground state can be anticipated.

The magnetic phase transitions induced
by applying a magnetic field in the low-dimensional magnets have attracted much interest recently from both experimental and theoretical points of view.  When the magnetic interactions cannot be satisfied simultaneously owing to the existing competing orders, the magnetic systems become fertile ground for the emergence of exotic states.
The ground-state magnetization plateaus appearing in  polymerized Heisenberg
chains under external magnetic fields was investigated, and the phase diagram of AFM bond alternating spin chain in homogeneous magnetic fields was presented~\cite{Honecker,Daniel}. Moreover, the effects of temperature, magnetic field and dimerized interaction on the spin and heat transport in dimerized Heisenberg chains in a magnetic field are studied. It is noted that the spin and heat conductivity show different behaviors in different phases~\cite{langer}. For alternating spin-1/2 chains with anisotropic AFM-FM coupling under a transverse magnetic field, two successive phase transitions, i.e.,  from Haldane phase to stripe AFM phase and from stripe AFM phase to polarized paramagnetic phase, have been identified to be Ising tpye~\cite{Saeed}.  The magnetization state and magnetic structure can be revealed through common techniques like neutron diffraction measurements and synchrotron X-ray scattering ~\cite{Matsuda,Matsuda2010}. However, the phase boundary of the spin-1/2 Heisenberg AFM-FM bond alternating chain in a magnetic field is still not clear, and needs to be discussed further. Fortunately, with the development of quantum information, various information measures, e.g., quantum coherence, entanglement entropy, and fidelity, can help us to study quantum critical phenomena in spin chains. It is found that the quantum critical points can be well characterized by both the ground-state entanglement and fidelity on large system~\cite{Ren,Chen,You07,Gu,You11,You15a,You15b,Ren02}. 
In this paper we study the entanglement, coherence and fidelity of the spin-1/2 Heisenberg alternating chain under a transverse magnetic field, and finally the phase diagram will be given.
\begin{figure}[t]
\includegraphics[width=0.50\textwidth]{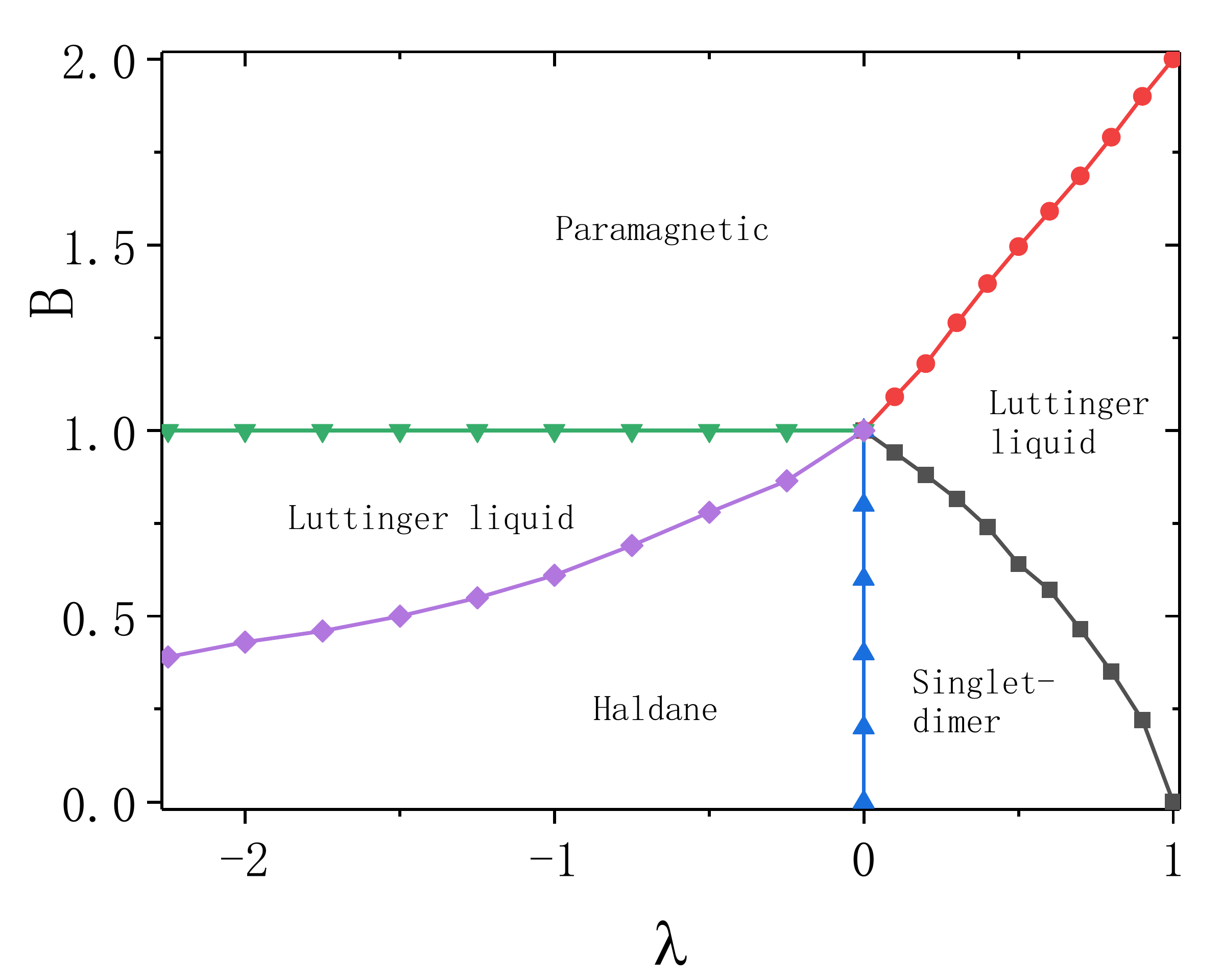}
\caption{ Phase diagram of spin-1/2 alternating Heisenberg chain ($\Delta$=1) as functions of the alternating interaction $\lambda$ and the magnetic field $B$.
\label{phasexxx}}
\end{figure}

\begin{figure*}[t]
\includegraphics[width=0.65\textwidth]{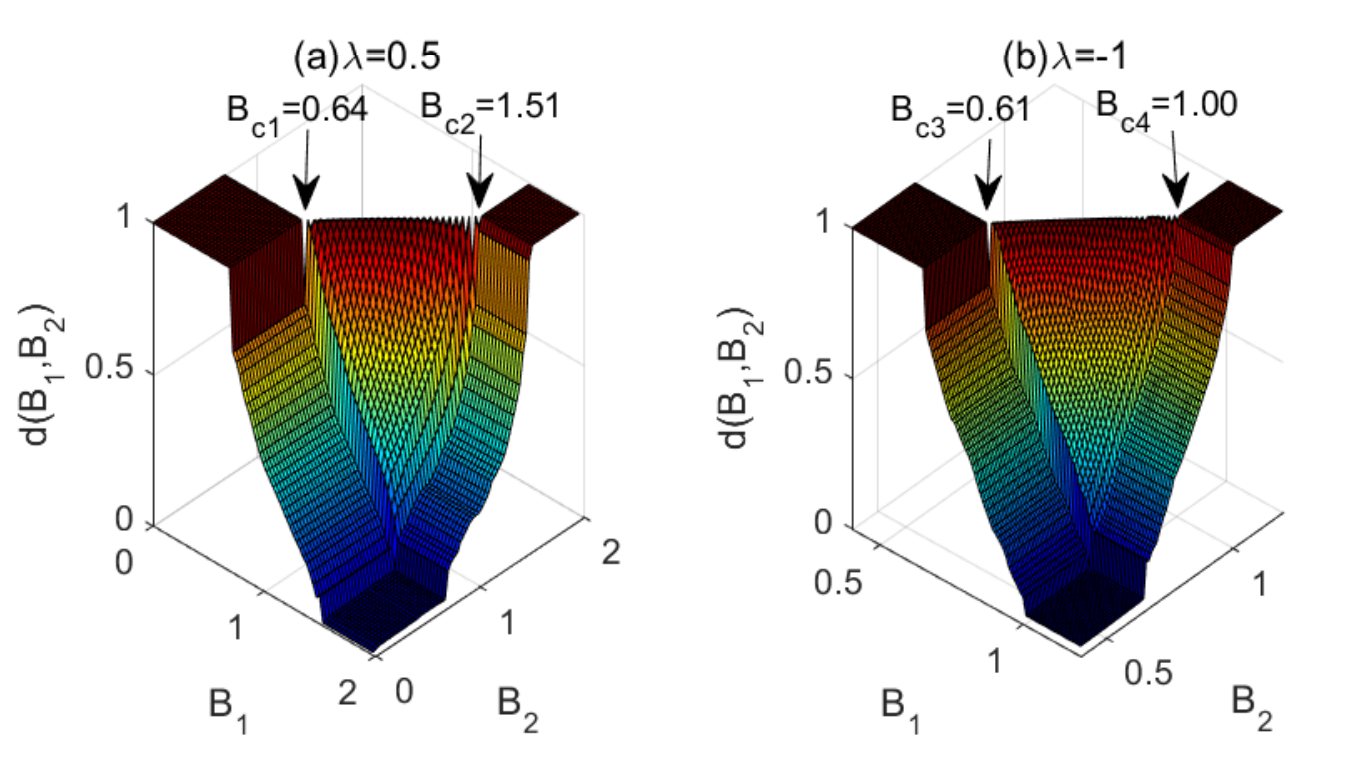}
\caption{Ground state fidelity per site $d(B_1,B_2)$ for the model with different alternating interactions (a)$\lambda=0.5$ (b)$\lambda=-1$.
\label{Fidelityxxx}}
\end{figure*}
\section{Hamiltonian and Measurements}
\label{sec:Hamiltonian}
The Hamiltonian of a one-dimensional (1D) spin chain is given by
\begin{equation}
\begin{split}
\label{eq1}
H=&\sum_{i=1}^{N/2}J(S_{2i-1}^x S_{2i}^x+S_{2i-1}^y S_{2i}^y+\Delta S_{2i-1}^z S_{2i}^z)\\
  &+\sum_{i=1}^{N/2}\lambda(S_{2i}^x S_{2i+1}^x+S_{2i}^y S_{2i+1}^y+\Delta S_{2i}^z S_{2i+1}^z)\\
&-B \sum_{i=1}^{N}S_i^z,
\end{split}
\end{equation}
where $S_i^\alpha(\alpha=x,y,z)$ are spin operators on the $i$-th site and $N$
is the length of the spin chain. The parameter $J$ is the AFM coupling on odd bonds, and $J$=1 is assumed hereafter in the paper.
$\lambda$ is considered to be either AFM ($\lambda > $0) or FM ($\lambda <$ 0) coupling strength on even bonds. The parameter $B$ is the strength of the magnetic field along the $z$-axis with the anisotropy $\Delta$. In the case of $\lambda=0$ the ground state is composed of local dimers on odd pairs $2j-1$ and $2j$. It is easy to obtain the energy per bond $E=1/4$, $-3/4$, $\Delta/4+B$, $\Delta/4-B$. As the increasing of $B$, the ground state of the system changes from a direct product of singlet pairs to the direct product of polarized qubits. 


For $\Delta=0$, Eq.(\ref{eq1}) reduces to the dimerized XX model, which can be solved by Jordan-Wigner transformation and Fourier-Bogoliubov
transformation.  It has been shown that
the dimerized XX model is equivalent to anisotropic XY model in given parity blocks~\cite{Venuti10}. It is more straightforward to see the equivalence between them in the fermionic form:
\begin{eqnarray}
H_{DXX}&=&\sum_{i=1}^{N/2}  \frac{1}{2} \left( c_{2i-1}^+c_{2i} + \lambda  c_{2i}^+c_{2i+1}+h.c.\right) \nonumber \\ &+&\sum_{i=1}^{N}\frac{B}{2}(1-2c_{i}^+c_{i})\}. \label{XX}
\end{eqnarray}
 Then, we use the local mapping under the assumption of even $N$ and periodic boundary condition:
\begin{eqnarray}
c_j^\dagger=\frac{1}{2}\left[ i a_{j+1}^\dagger +a_{j}^\dagger - (-1)^j (i a_{j+1} + a_j) \right] ,
\end{eqnarray}
and thus Eq.(\ref{XX}) can be transformed into a generalized anisotropic model:
\begin{eqnarray}
H_{AXX}&=&\sum_{j=1}^{N}   \left( J_h a_{j}^\dagger a_{j+1} + J_p  a_{j}^\dagger a_{j+1}^\dagger+h.c.\right) \nonumber \\
&+&\sum_{j=1}^{N} \frac{iB}{2}\left[a_j^\dagger a_{j+1} -(-1)^j  a_j^\dagger a_{j+1}^\dagger  +h.c. \right], \label{AXX}
\end{eqnarray}
where $J_h$=$(1+\lambda)/4$, $J_p$=$(\lambda-1)/4$.
Eq. (\ref{AXX}) can be traced back to the spin version:
\begin{eqnarray}
\label{eq3}
H
&=&\sum_{j=1}^{N} (\lambda S_{j}^x S_{j+1}^x+   S_{j}^y S_{j+1}^y)\nonumber \\
&+&B\sum_{j=1}^{N/2}(S_{2j-1}^y S_{2j}^x- S_{2j}^x S_{2j+1}^y).
\end{eqnarray}
 For $B=0$ the QPT occuring at $\lambda=1$ for the dimerized chain shares the same properties with the transition which occurs at $\lambda=1$ for the anisotropic XY chain separating the $x$-component phase from the $y$-component N\'{e}el phase.
In the case of
$\lambda=1$ dimerized XX model corresponds to the uniform XX chain. In the opposite limit $\lambda=0$ one
arrives at a collection of isolated (uncoupled) XX dimers.
$B$ term in Eq. (\ref{eq3}) favors period-4 configurations $\vert \uparrow \leftarrow \downarrow \rightarrow \cdots \rangle$ or $\vert  \downarrow \rightarrow \uparrow \leftarrow \cdots\rangle$, competing with  $x$-component and $y$-component N\'{e}el orderings.
The generic features of the spin-Peierls systems are analytically discussed in detail
by Taylor and M\"{u}ller \cite{Taylor85}. The details can be referred to Appendix \ref{appendix-XX}.


As we know, it is difficult to diagonalize the Hamiltonian Eq.(\ref{eq1}) when $\Delta \neq 0$.
The finite-size density matrix renormalization group~(DMRG) would be the effective method to obtain the ground-state wavefunctions approximately~\cite{white,U01}. In this version of DMRG, an open chain is grown iteratively by adding two sites at a time to the center of the spin chain, and up to the sizes $N=400$. Then, we perform four sweepings, and the maximum number of the eigenstates kept is $m=200$ during the processing. 
Such truncation guarantees that the converging error is smaller than $10^{-7}$. With this accurate calculation, we can precisely analyze the QPTs through various theoretic measures.

A QPT taking place in this class of systems has been thoroughly investigated in the thermodynamic limit. However, both experimental and theoretical difficulties have boosted a high interest in finite-size systems, which show the “forerunners” of the points of QPT of the
thermodynamic systems.  In general, finite-size systems would exhibit many energy level crossings between
physical and unphysical states.  As a consequence, diverse theoretic measures would have some jumps.
In order to avoid the finite-size effects, we also implement the infinite time-evolving block decimation (iTEBD)~\cite{vidal01,vidal02}, which can be used to compute the ground-state wavefunctions for an infinite-size lattice in one or two dimensions with translational invariance.  It can help us directly address physical quantities in the thermodynamic limit with high quality. Given a large bond dimensions $\chi$, the ground-state wavefunctions based on the matrix product state representations can be obtained by applying imaginary-time evolution gates $\exp(-\tau h)$ on a given initial random state $\vert \psi(0)\rangle$, until the latter converges to the variational ground state. Here $h$ is the local Hamiltonian, which is composed of two-site coupling terms on an odd bonds $h_{2i-1,2i}$ or even bonds $h_{2i,2i+1}$, and $\tau$ is the Trotter step length. In practice,
we start from $\tau = 0.1$ and gradually reduce it by $\tau=\tau/10$, and break the loop until $\tau< 10^{-9}$.
In the paper, $\chi=50$ is adopted, and we check our codes with the case
$\lambda=1$, $B=0$, which is equal to well-known Heisenberg chain. The ground-state energy we obtain is $E_0=-0.443143049$, which is very close to the exact diagonalization result $E=1/4-\ln(2)=-0.4431471805$, and the error is smaller than $5.0\times10^{-6}$.

As the external parameter varies across a critical point, the ground-state wavefunction undergoes a sudden change in the wake of QPT, accompanied by a rapid alteration in a variety of quantum measurements. Fidelity is one of the most effective measurements, which can detect the
 critical points~\cite{zhou,Wang,Zhou2018}. It measures the 
 similarity between the two closest states as the external parameter such as $\kappa$ is tuned, which is defined as
$F_N(\kappa_1,\kappa_2)=|\langle\psi(\kappa_1)|\psi(\kappa_2)\rangle|$ in finite-size systems. In the thermodynamic limit, the ground-state fidelity per site
\begin{equation}
d(\kappa_1,\kappa_2)=\lim_{N\rightarrow\infty}F_N^{1/N}(\kappa_1,\kappa_2)
\end{equation}
can be calculated easily in iTEBD by transfer matrix~\cite{Zhou2018}.
The fidelity $d(\kappa_1,\kappa_2)$ should be equal to one when $\kappa_1=\kappa_2$, and QPTs may be detected through singularities exhibited in $d(\kappa_1,\kappa_2)$. Meanwhile, the concurrence is chosen as a measure of the
pairwise entanglement between two qubits~\cite{Wootters}. The concurrence $C$ is
defined as
\begin{equation}
\label{eq2} C(\rho_{12})= \max \{{\beta_1 - \beta_2 - \beta_3 -
\beta_4 ,0}\},
\end{equation}
where the quantities $\beta_i (i=1, 2, 3, 4)$ are the square roots
of the eigenvalues of the operator $\varrho = \rho_{12}(\sigma_1^y
\otimes \sigma_2^y)\rho_{12}^\ast (\sigma_1^y \otimes \sigma_2^y)$ and in descending order.
The case of $C=1$ corresponds to the
maximum entanglement between the two qubits, while $C=0$ means that
there is no entanglement between the two qubits. The entanglement entropy
is used as a measure of the bipartite entanglement. If
$|\psi\rangle$ is the ground state of a chain of $N$ qubits, a reduced
density matrix of contiguous qubits from $1$ to $L$ can be written as
$\rho_L=\textrm{Tr}_{N-L}|\psi\rangle\langle \psi|$.
The bipartite entanglement between the right-hand $L$ contiguous
qubits and the rest of the system can be measured by the entropy
\begin{equation}
\label{eqSL}S_L(\rho_{1\cdots L})=-{\rm Tr}(\rho_L\log_2\rho_L).
\end{equation}
One of the basic properties of the block entanglement entropy for a pure state can be
given by $\label{eq5}S_L=S_{N-L}$,
since the spectrum of the reduced density matrix $\rho_L$ is the
same as that of $\rho_{N-L}$ following from the Schmidt decomposition. This property implies the entanglement entropy
is not extensive and boils down to a celebrated area law for non-critical ground states or finite temperature system, which states
the leading term of the entanglement entropy is proportional to the boundary area between $L$ and $N-L$ qubits. Note that
the area law is violated for highly excited states and for ground state of gapless systems.

\section{Results}
\label{sec:results}

\begin{figure}[t]
\includegraphics[width=0.51\textwidth]{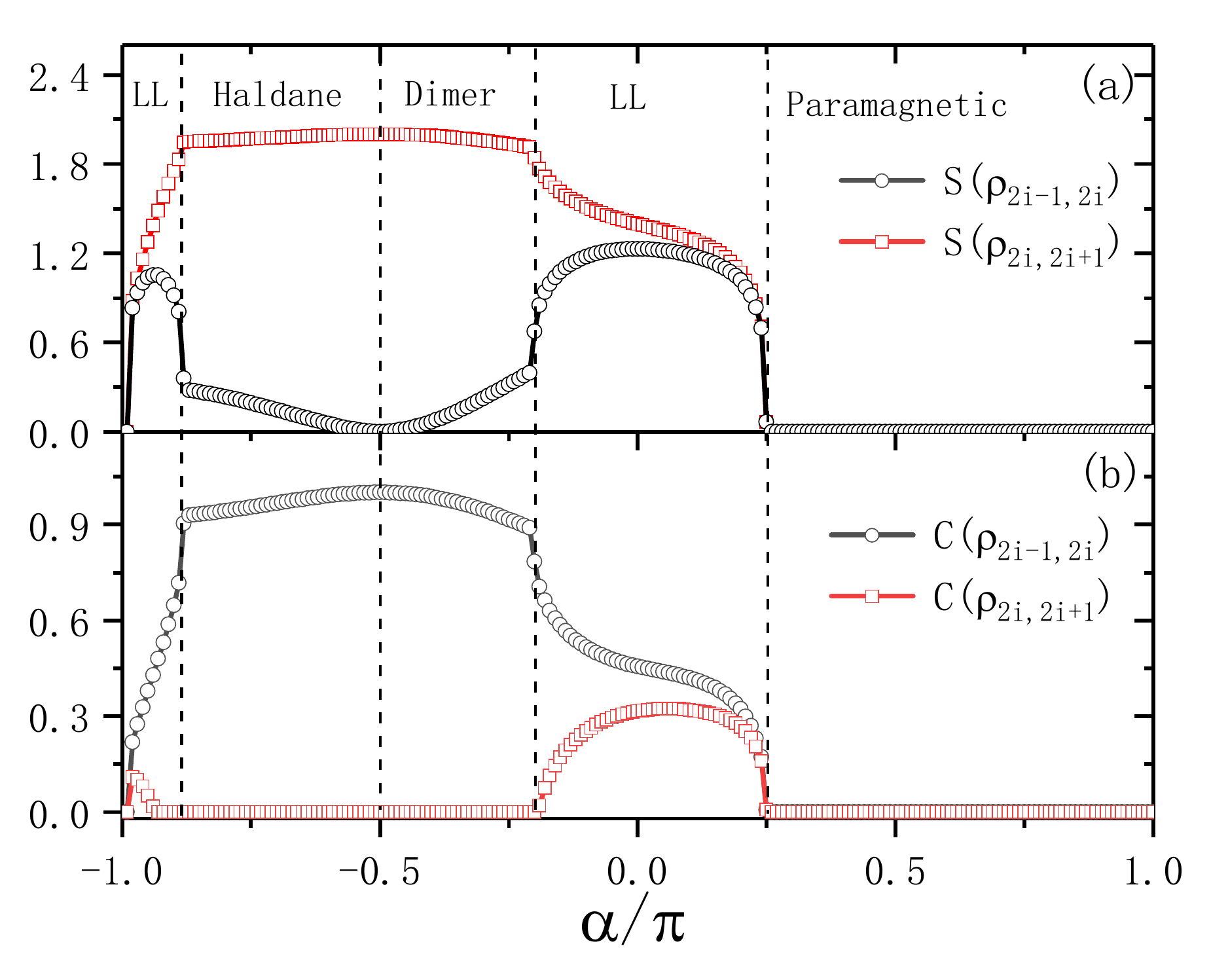}
\caption{(a) Entanglement entropy and (b) concurrence are plotted as function of $\alpha$ with $R=0.8$. Here LL denotes the Luttinger liquid phase.
\label{entanglement_XXX}}
\end{figure}
\begin{figure}[t]
\includegraphics[width=0.50\textwidth]{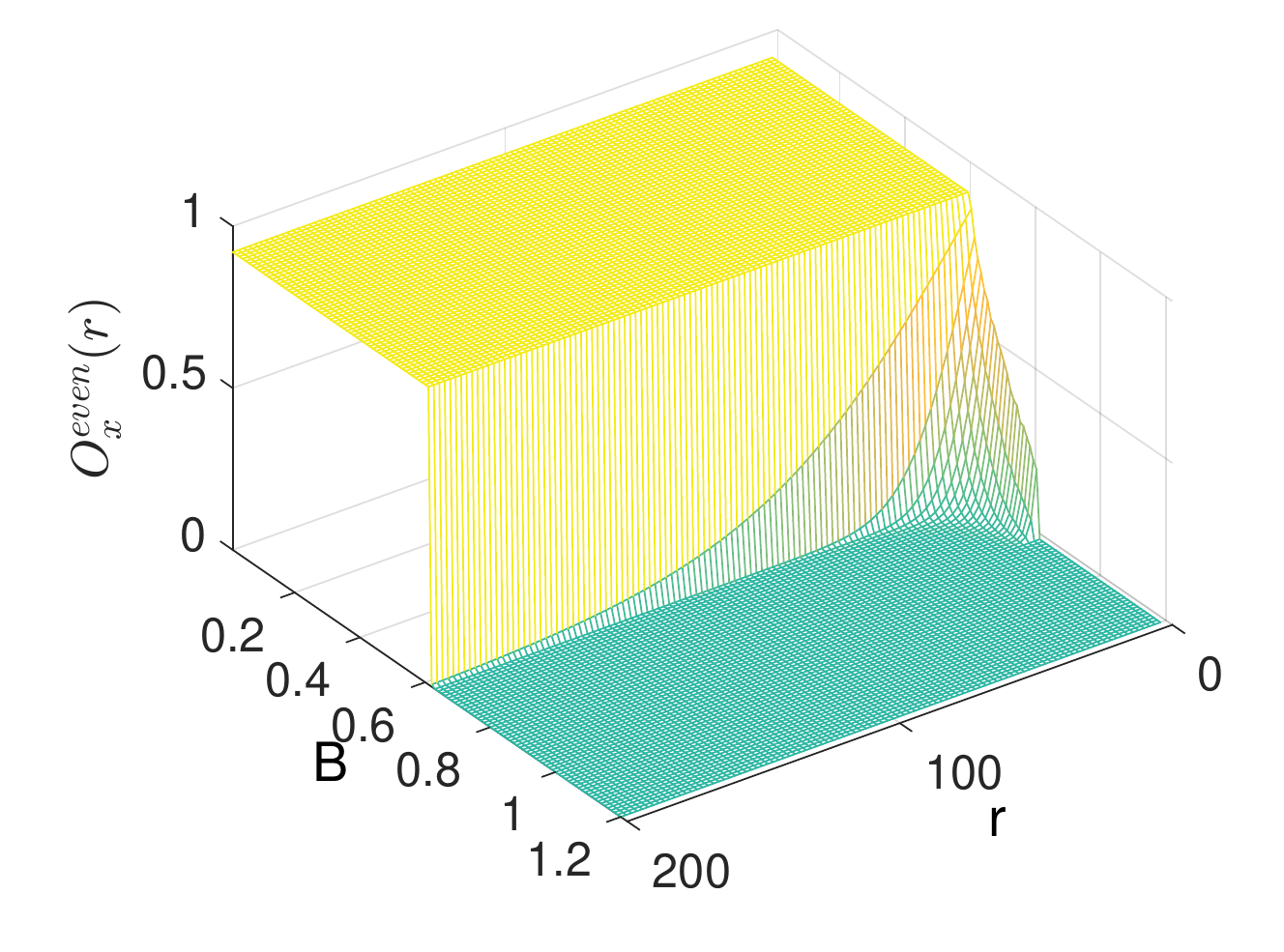}
\caption{The string order parameter $O^{even}_x$ ($r$) is plotted as a function of
the magnetic field $B$ and site distance $r$ for with $\lambda=-1$.
\label{SOP_XXX}}
\end{figure}

The ground-state phase diagram of the Hamiltonian Eq. (\ref{eq1}) for $\Delta=1$ is shown in Fig. \ref{phasexxx}. Note that the phase diagram on $B$-$\lambda$ plane has been shown in Refs.\cite{Honecker,langer} for $\lambda>0$.  We calculated the fidelity per site to check our codes, which is shown in Fig. \ref{Fidelityxxx}(a). It is seen that $d(B_1,B_2)=1$ when $|\psi(B_1)\rangle,|\psi(B_2)\rangle$ are both in the dimerized phase or both in the paramagnetic phase at the same time.  The two pinch points which mean the critical points can be identified as $B_{c1}=0.64$, $B_{c2}=1.53$. The values agree with the results obtained by the spin gap~\cite{Uhrig,Yu,Chitra}. When $B_{c1}<B<B_{c2}$, the model will be in the Luttinger liquid phase, which is gapless. We also show the phase diagram for $\lambda<0$. It is obvious that the model would be in Haldane phase for $B=0$ and paramagnetic phase with large magnetic field. The Luttinger liquid phase separates these phases for intermediate $B$~\cite{sakai}. The critical points between them can also be
portrayed by the fidelity. The result is shown in Fig. \ref{Fidelityxxx}(b). The two pinch points can be found $B_{c3}=0.61$, $B_{c4}=1.00$ for $\lambda=-1$.

Interestingly, all the critical lines cross at one point $B=1$, $\lambda=0$, which is a multicritical point. In order to better compare various measures, we inspect the QPTs along a loop path described by the radius $R$ and the angle $\alpha$:
 \begin{equation}
 \begin{split}
 &B=(1+\Delta)/2+R \sin\alpha,\\
 &\lambda=R \cos\alpha,
 \end{split}
 \end{equation}
where $\alpha$ ranges from $\alpha=-\pi$ to $\alpha=\pi$.
The entanglement entropy between two qubits and the rest of the system is shown in Fig.\ref{entanglement_XXX} with $R=0.8$.
When $\alpha/\pi\simeq -1$, the system is in the Luttinger liquid phase. With the increasing of $\alpha$, the system will be in the Haldane phase, and the critical point is uncovered by the discontinuity of entanglement entropy of even or odd bond. When $\alpha$ increases further, the entanglement entropy of even  bond $S_{2i,2i+1}$ will rise, and entanglement entropy of odd bond $S_{2i-1,2i}$ will decline. When $\alpha=-\pi/2$, $S_{2i,2i+1}=1$ and $S_{2i,2i+1}=0$, which pinpoint the phase transition point between the Haldane phase and the dimer phase. For the Haldane phase, the system has $|\uparrow\downarrow\downarrow\uparrow\rangle_{1,2,3,4}\cdots\otimes|\uparrow\downarrow\downarrow\uparrow\rangle_{2i-1,2i,2i+1,2i+2}\cdots$, and $|\uparrow\downarrow\uparrow\downarrow\rangle_{1,2,3,4}\cdots\otimes|\uparrow\downarrow\uparrow \downarrow\rangle_{2i-1,2i,2i+1,2i+2}\cdots$ for singlet dimer. The singlet dimer-Luttinger liquid  and Luttinger liquid- paramagnetic phase transitions can be captured by discontinuity of the entanglement entropy. It is noted that the Haldane-singlet dimer can be detected by either the peak of $S(\rho_{2i,2i+1})$ or the valley of $S(\rho_{2i-1,2i})$. This is because at this moment the odd bond is one of Bell states, and the concurrence of odd bond reaches the maximal value 1, which is shown in Fig. $\ref{entanglement_XXX}$(b), so the entanglement between the odd bond and the rest of the system measured by entanglement entropy will go to zero. The counterpart for the even bond is opposite. Significantly, the concurrence for the even bond can not detect the Luttinger liquid-Haldane and Haldane-singlet dimer phase transition.

\begin{figure}[t]
\includegraphics[width=0.5\textwidth]{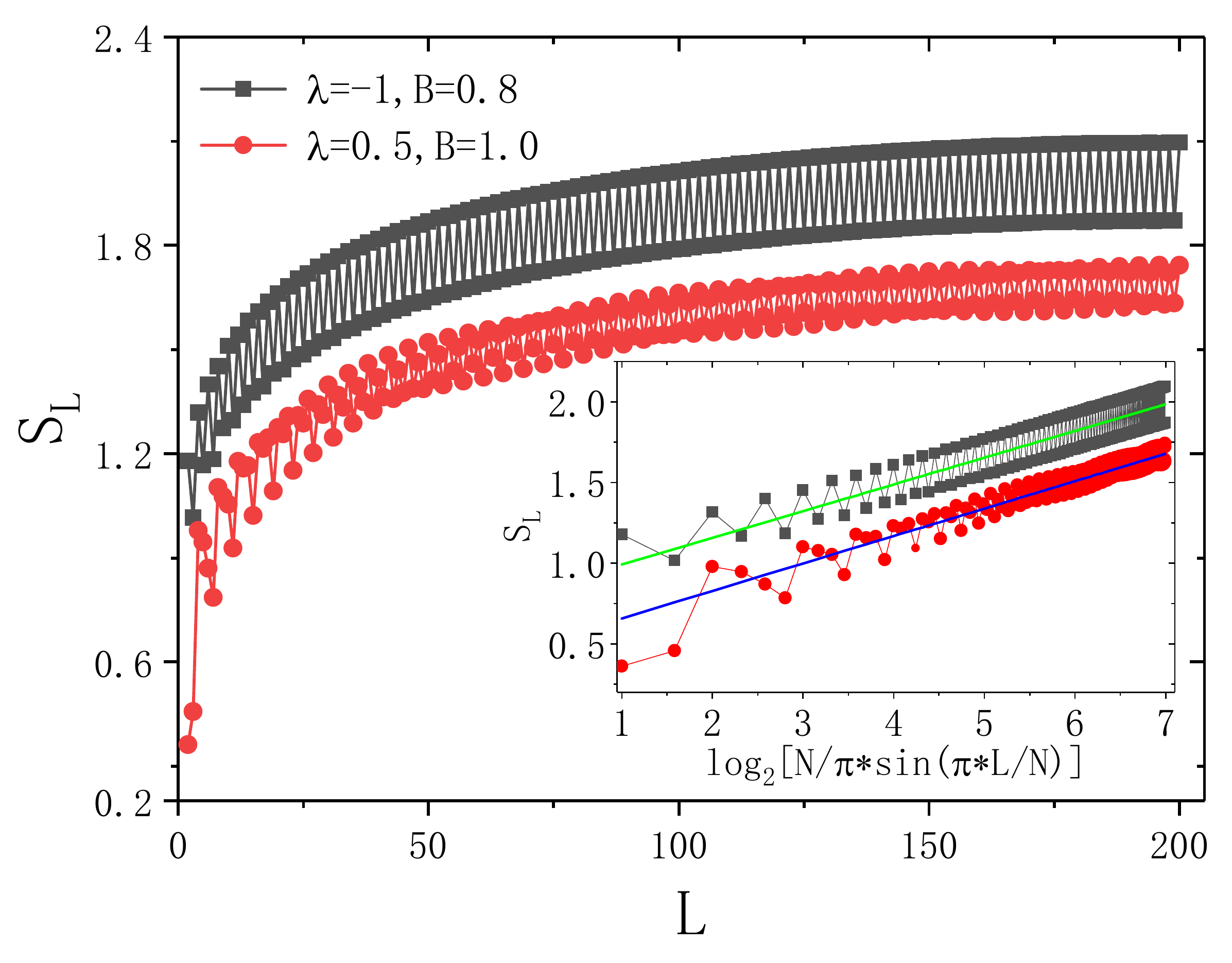}
\caption{Entanglement entropy $S_L$ between contiguous $L$ qubits and the
remaining $N-L$ qubits is plotted as a function of $L$ and $\log_2[\frac{N}{\pi} \sin(\frac{\pi L}{N})]$ (inset) for $N=400$. The lines are the numerical fittings.
\label{EE_scaling_XXX}}
\end{figure}

\begin{figure}[t]
\includegraphics[width=0.5\textwidth]{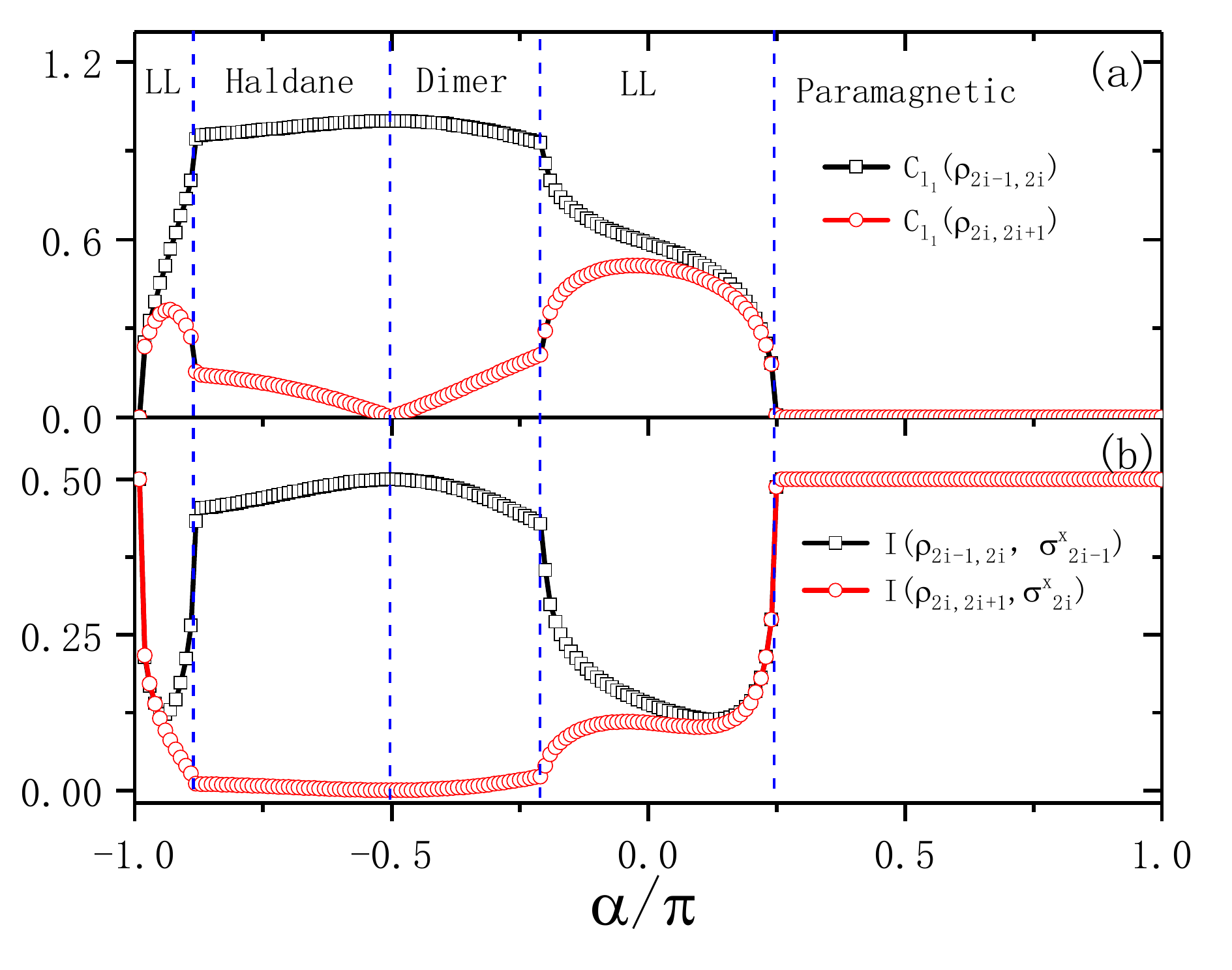}
\caption{(a) Coherence $C_{l_{1}}$ (b) the Wigner-Yanase skew information $I(\rho,\sigma^x)$ are plotted as function of $\alpha$ with $R=0.8$. Here LL denotes the Luttinger liquid phase.
\label{coherence_XXX}}
\end{figure}

We also calculate the string order parameter (SOP)\cite{Hida1993,Lou,Liu01}.
The SOP characterizes the topological order in the Haldane phase of $S=1$ Heisenberg chain efficiently~\cite{Ren03}. It is noted that the SOP will behave as an oscillation in dimerized model~\cite{Liu,Paul}. In the paper, we adopt the version of SOP in terms of
$S=1/2$ operators, which is given by
\begin{equation}
O^{even}_x(r) = -4\langle S_i^{x} \exp(i \pi \sum_{i<l<j} S_l^{x}) S_{j}^{x}\rangle.
\label{eq9}
\end{equation}
Here we set $i$ = 1 and $j$ being an even number, and thus the distance $r=|i-j|$  is an odd integer.
The results are presented in Fig. \ref{SOP_XXX}. It is seen that the SOP is almost one, and has little change with the distance $r$ when $0<B<0.61$. We can surmise that the SOP is not equal to zero when $r\rightarrow\infty$.
 As long as $B$ increases further, the SOP decays suddenly, indicating the system changes to Luttinger liquid  phase. When $r$ is small, the SOP has a very small value, and decreases to zero rapidly, because the short-range correlations in Luttinger liquid phase are large and decreases exponentially with the increasing distance. When the system in the PM phase, the SOP is equal to zero for any $r$.

The entropy $S_L$ between contiguous $L$ qubits and the
remaining $N-L$ qubits in the Luttinger liquid phase is plotted as a function of the subsystem length $L$ for $N=400$ in Fig. \ref{EE_scaling_XXX}.
One finds there are large odd-even effects for $\lambda=-1.0,B=0.8$.
When $\lambda=0.5,B=1$, the oscillations become more complicated. Such phenomenon is caused by the open boundary condition and bond alternation.
The entanglement entropy $S_L$ in the critical regime swells up with subsystem size $L$($L\leq N/2$), which is
predicted by conformal field theory~(CFT) as
\begin{equation}
\label{eq6}S_L \sim \frac{c}{6}\log_2[\frac{N}{\pi}\sin(\frac{\pi}{N}L)]+A,
\end{equation}
where $c$ is the central charge and $A$ is a non-universal constant~\cite{Calabrese,Chiara,Nicolas,Ren01}.  The entropy $S_L$ is also plotted as
a function of $\log_2[\frac{N}{\pi}\sin(\frac{\pi}{N}L)]$ in the inset of Fig. \ref{EE_scaling_XXX}. It is shown that the entropy appears as a straight line, whose slope
yields  
$c\simeq 1.0$ for both cases~\cite{sakai}. 
Such a multiplicative logarithmic scaling is known to violate the celebrated area law, which states the entanglement entropy between two subsystems scales with the boundary between them. As for non-critical ground states of one dimensional quantum systems, the area law yields a constant entanglement entropy, independent of the subsystem size.

Recently, quantum coherence is a resurgent concept in the quantum theory and acts as a manifestation of the quantum superposition principle. We also recall the $l_1$ norm of coherence~\cite{Plenio}, which can almost be estimated in experiments~\cite{Zhang,Zheng}.
For a density matrix $\rho$ in the reference
basis $\{|i\rangle\}$, $l_1$ norm of coherence is given by
\begin{equation}
C_{l_1}(\rho)=\sum_{i\neq j}\langle i|\rho|j\rangle.
\label{eq2}
\end{equation}
Moreover, the local quantum coherence and the local quantum uncertainty, based on Wigner-Yanase skew information (WYSI), given by \cite{Girolami}
\begin{equation}
I(\rho,K)=-\frac{1}{4}\textrm{Tr}([\rho,K]^2),
\label{eq4}
\end{equation}
where the density matrix $\rho$ describes a quantum state, $K$ plays a role of
an observable, and [.,.] denotes the commutator. The WSYI shows the capability of diagnosing the QPT in the anisotropic XY chain~\cite{Karpat}. Similar with the entanglement entropy, the quantum coherence is shown in Fig. \ref{coherence_XXX}. All the phase transitions can be identified by divergences or discontinuity of coherence for even bonds and odd bonds.

\section{Conclusions }
\label{sec:Discussion}
By using the infinite time-evolving block decimation, the ground-state properties of the spin-1/2 Heisenberg alternating chain in a magnetic field are calculated with very high accuracy. We numerically investigate the effects of magnetic field on the fidelity, which measures the similarity between two states, and then analyze its relation with the quantum phase transitions (QPTs). QPTs are intuitively accompanied by an abrupt change in the structure of the ground-state wave function, so generally a pinch point
of the fidelity indicates a QPT and the location of the pinch point denotes the critical point. Based on the above analyisis, we obtain the phase diagram. This model supports the Haldane phase, the singlet-dimer phase, the Luttinger liquid and the paramagnetic phase.
In the Luttinger liquid phase, we study the scaling of the entanglement entropy with the subsystem size $L$, and identified the central charge $c=1$. We also study the quantum coherence, whose anomalies detect all the phase transitions therein. In summary, conclusions drawn from these quantum information observables agree well with each other.

\vspace{0.5cm}
\begin{acknowledgments}

This work is supported by the National Natural Science
Foundation of China (NSFC) (Grants Nos.11384012, 11474211). J. R thanks Juan Felipe Carrasquilla for his discussion with iTEBD codes.

\end{acknowledgments}

\vspace{0.5cm}
\begin{appendix}
\section{Exact solution of dimerized XX model}
\label{appendix-XX}

Consider a chain of spin-1/2 operators interacting antiferromagnetically with their
nearest neighbors, given by
\begin{eqnarray}
H&=&\sum_{i=1}^{N/2}J(S_{2i-1}^x S_{2i}^x+S_{2i-1}^y S_{2i}^y )+ \lambda(S_{2i}^x S_{2i+1}^x+S_{2i}^y S_{2i+1}^y )
   \nonumber \\
   &-&B \sum_{i=1}^{N}S_i^z. \label{Hxx}
\end{eqnarray}
The dimerized XX model corresponds to $\Delta$=0 in Eq. (\ref{eq1}).
The Hamiltonian (\ref{Hxx}) can be exactly diagonalized following
the standard procedure for 1D systems. In terms of the raising and lowering operators for spins,
\begin{equation}
\begin{split}
   S_i^+=S_i^x+iS_i^y, \quad S_i^-=S_i^x-iS_i^y,
          \end{split}
    \end{equation}
The Hamiltonian (\ref{Hxx}) then takes the form:
\begin{eqnarray}
 H&=&\sum_i\frac{1}{2} (S_{2i-1}^+S_{2i}^-+S_{2i-1}^-S_{2i}^+)+ \frac{\lambda}{2}(S_{2i}^+S_{2i+1}^-+S_{2i}^-S_{2i}^+)\nonumber
 \\ &+&BS_i^z .
    \end{eqnarray}
The Jordan-Wigner transformation
maps explicitly between quasispin operators and spinless fermion
operators by
\begin{eqnarray}
S _{j}^{+}& =&\exp \left[ i \pi \sum_{i=1}^{j-1}c_{i}^{\dagger }c_{i}%
\right] c_{j}=\prod_{i=1}^{j-1}(2S_{i}^{z})c_{j},  \notag \\
S _{j}^{-}& =&\exp \left[ -i\pi \sum_{i=1}^{j-1}c_{i}^{\dagger }c_{i}%
\right] c_{j}^{\dagger }=\prod_{i=1}^{j-1}(2 S
_{i}^{z})c_{j}^{\dagger },
\notag \\
S _{j}^{z}& =&1/2-c_{j}^{\dagger }c_{j}, \label{JW}
\end{eqnarray}
where $c_{j}$ and $c_{j}^{\dagger }$ are annihilation and creation
operators of spinless fermions at site $j$, which obey the standard
anticommutation relations, $\{c_{i}, c_{j}\}=0$, $\{c_{i}^{\dagger
}, c_{j}\}=\delta_{ij}$. By substituting Eq. (\ref{JW}) into Eq. (\ref{Hxx}), we find
\begin{eqnarray}
         H&=&\sum_{i=1}^{N/2} \{\frac{1}{2}(c_{2i-1}^+c_{2i}+c_{2i-1}c_{2i+1}^+) +\frac{\lambda}{2}(c_{2i}^+c_{2i+1}+c_{2i}c_{2i+1}^+)\nonumber\\
        &+&\sum_{i=1}^{N}\frac{B}{2}(1-2c_{i}^+c_{i})\}.
\end{eqnarray}
Next, utilizing translational invariance, discrete Fourier transformation for plural spin sites is
introduced by
\begin{eqnarray}
c_{2j-1}=\sqrt{\frac{2}{N }}\sum_{k}e^{-ik j}a_{k},\text{ \ \ }c_{2j}=%
 \sqrt{\frac{2}{N }}\sum_{k}e^{-ik j}b_{k},\text{ \ \ }
\end{eqnarray}
with the discrete momentums as
\begin{eqnarray}
k=\frac{2n\pi}{ N },  n= -(\frac{N}{2}-1), -(\frac{N}{2}-3),
\ldots, \frac{N}{2}-1. \;
\end{eqnarray}

\begin{figure}[t]
\includegraphics[width=0.450\textwidth]{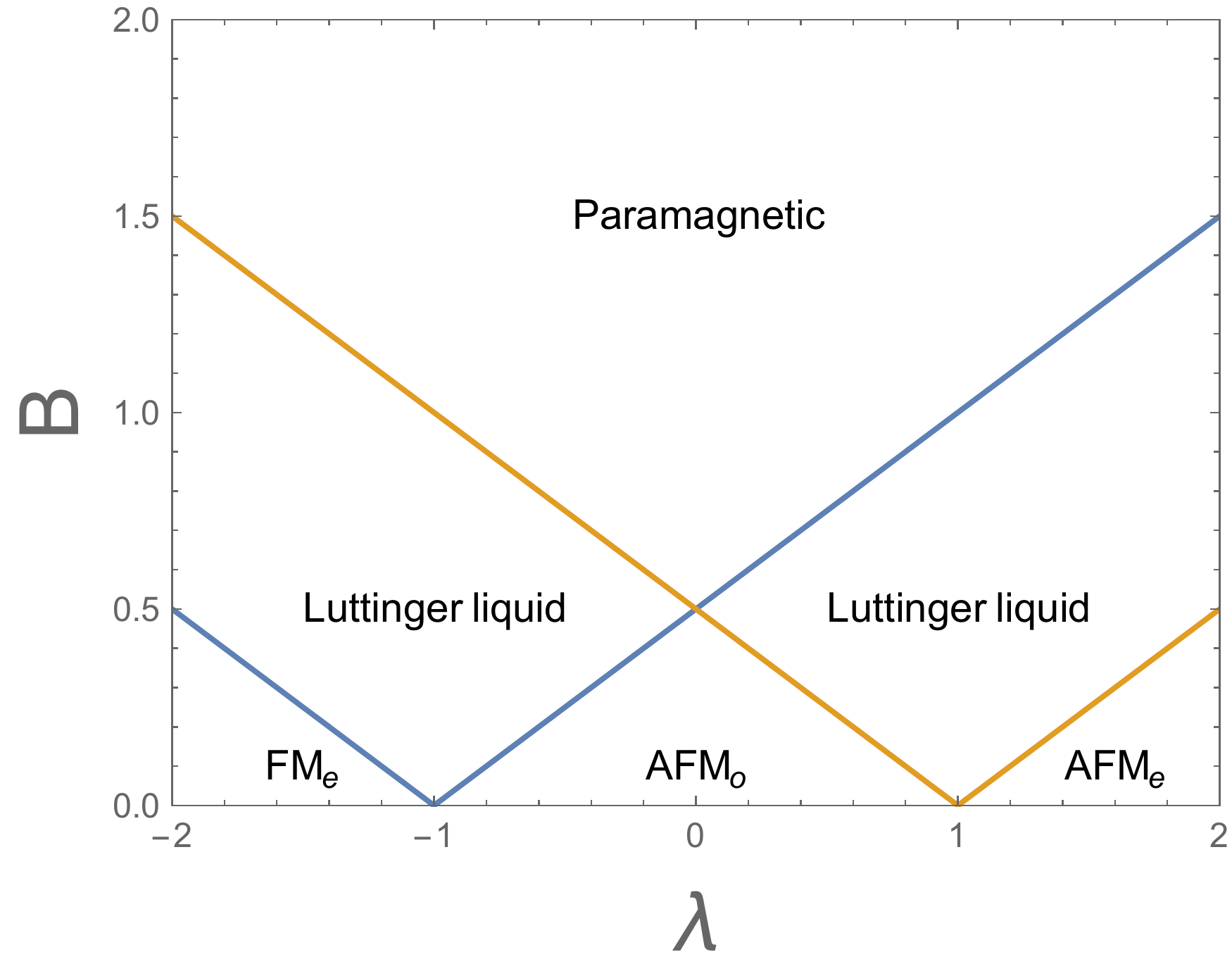}
\caption{ Phase diagram of spin-1/2 alternating XX chain as functions of the alternating interaction $\lambda$ and the magnetic field $B$.
\label{phasexx}}
\end{figure}
Let's proceed with diagonalization of the dimerized XX model. We rewrite,
\begin{eqnarray}
      & H=\sum_{k}(-\frac{1}{2}-\frac{\lambda}{2}e^{ik})b_ka_k^+ + \sum_{k}(\frac{1}{2}+\frac{\lambda}{2}e^{ik})b_k^+a_k\nonumber \\
      &+\sum_{k}\frac{B}{2}(1-2a_k^+a_k)+\sum_{k}\frac{B}{2}(1-2b_k^+b_k).\label{Hxxk}
\end{eqnarray}
    Defining
\begin{equation}
\begin{split}
      &\Lambda_k=\frac{1}{2}+\frac{\lambda}{2}e^{ik},
          \end{split}
    \end{equation}
then Eq.(\ref{Hxxk}) can be rewritten as
\begin{eqnarray}
      &H=\sum_k(\Lambda_k a_k^+b_k+\Lambda^*_kb_k^+a_k)+\sum_kB(1-a^+_ka_k-b_k^+b_k).\nonumber \\
\end{eqnarray}
 Then we
write the Hamiltonian in the matrix form:
\begin{equation}
H=\sum_k
 \left(                 
  \begin{array}{cc}   
    a_k^+ & b_k^+ \\  
  \end{array}
\right)        
\left(                 
  \begin{array}{cc}   
    -B & \Lambda_k \\  
   \Lambda_k^* & -B \\  
  \end{array}
\right)
    \left(                 
  \begin{array}{cc}   
    &a_k \\
    & b_k   
  \end{array}
\right)
 +\sum_k B.              
\end{equation}
The eigenspectra can be obtained as:
\begin{equation}
\begin{split}
&\varepsilon_{\pm,k}=\pm|\Lambda_k|-B.\\
    \end{split}
    \end{equation}
Consequently, the Hamiltonian can be written in the diagonal form:
\begin{eqnarray}
  H=\sum_k[\varepsilon_{+,k}(\gamma_{1,k}^+\gamma_{1,k}-\frac{1}{2})+\varepsilon_{-,k}(\gamma_{2,k}^+\gamma_{2,k}-\frac{1}{2})],\nonumber \\
\end{eqnarray}
where
\begin{eqnarray}
&&\varepsilon_{+,k}=\frac{1}{2}\sqrt{1+\lambda^2+2\lambda \cos k}-B,\nonumber\\
&&\varepsilon_{-,k}=-\frac{1}{2}\sqrt{1+\lambda^2+2\lambda \cos k}-B. \nonumber
\end{eqnarray}
As a magnetic field is turned on, the one-particle
spectrum will simply moves to higher energies with its shape unchanged.
Since the $\varepsilon_{-,k}$ is always negative, the gap closing can be identified by
$ \partial \varepsilon_{+,k}/\partial k =0$,
which suggests that the boundaries are described by $B$=$|1+\lambda|/2$ for $k=0$ and  $B$=$|1-\lambda|/2$ for $k=\pi$ as shown in Fig.\ref{phasexx}. 

\end{appendix}


\begin{thebibliography}{25}

\bibitem{Sachdev} S. Sachdev, \emph{Quantum Phase Transitions} (Cambridge University Press, Cambridge, England, 1999).

\bibitem{Yu} W. Yu and S. Haas, Excitation spectra of structurally dimerized and spin-Peierls chains in a magnetic field, Phys. Rev. B \textbf{62}, 344 (2000).
\bibitem{FHM} F. Heidrich-Meisner, A. Honecker, D. C. Cabra, and W. Brenig, Zero-frequency transport properties of one-dimensional spin-$1/2$ systems, Phys. Rev. B \textbf{68}, 134436 (2003).

\bibitem{Orignac} E. Orignac, R. Chitra, and R. Citro, Thermal transport in one-dimensional spin gap systems, Phys. Rev. B \textbf{67}, 134426
(2003).

\bibitem{Tennant} D. A. Tennant, C. Broholm, D. H. Reich, S. E. Nagler, G. E. Granroth, T. Barnes, K. Damle, G. Xu, Y. Chen, and B. C. Sales, Neutron scattering study of two-magnon states in the quantum magnet copper nitrate, Phys. Rev. B \textbf{67}, 054414 (2003).

\bibitem{James}A. J. A. James, F. H. L. Essler, R. M. Konik, Finite Temperature Dynamical Structure Factor of Alternating Heisenberg Chains, Phys. Rev. B \textbf{78}, 094411(2008).

\bibitem{Klyushina}E. S. Klyushina, A. C. Tiegel, B. Fauseweh,  A. T. M. N. Islam,  J. T. Park, B. Klemke, A. Honecker, G. S. Uhrig, S. R. Manmana, and B. Lake,  Magnetic excitations in the S=1/2 antiferromagnetic-ferromagnetic chain compound BaCu$_2$V$_2$O$_8$ at zero and finite temperature, Phys. Rev. B \textbf{93}, 241109(R) (2016).


\bibitem{xiong} H. N. Xiong, J. Ma, Z. Sun, X. Wang, Reduced-fidelity approach for quantum phase transitions in spin-$1/2$ dimerized Heisenberg chains, Phys. Rev. B, \textbf{79}, 174425 (2009).

\bibitem{HTW} H. T. Wang, B. Li, S.Y. Cho, Topological quantum phase transition in bond-alternating spin-$1/2$ Heisenberg chains, Phys. Rev. B, \textbf{87}, 054402 (2013).

\bibitem{Okamoto17} Yoshihiko Okamoto, Daisuke Nakamura, Atsushi Miyake, Shojiro Takeyama, Masashi Tokunaga, Akira Matsuo, Koichi Kindo, and Zenji Hiroi, Phys. Rev. B \textbf{95}, 134438(2017).

\bibitem{Xu}Hong-Ze Xu, Shun-Yao Zhang, Guang-Can Guo, and Ming Gong, Exact dimerized phase in anisotropic XYZ model for quasi-one-dimensional magnets, arXiv:1806.05814.

\bibitem{Bulaevskii} L. N. Bulaevskii, Theory of Non-uniform Antiferromagnetic Spin Chains, JETP, \textbf{17}, 684 (1963).

\bibitem{Kohmoto} M. Kohmoto, H. Tasaki, Hidden Z$_2\times$Z$_2$ symmetry breaking and the Haldane phase in the $S=1/2$ quantum spin chain with bond alternation, Phys. Rev. B, \textbf{46}, 3486, (1992).

\bibitem{Hida}K. Hida, Crossover between the Haldane-gap phase and the dimer phase in the spin$-1/2$ alternating Heisenberg chain, Phys. Rev. B, \textbf{45}, 2207 (1992).

\bibitem{Hase} M. Hase, I. Terasaki, K. Uchinokura, Observation of the spin-Peierls transition in linear Cu$^{2+}$ (spin$-1/2$) chains in an inorganic compound CuGeO$_3$, Phys. Rev. Lett. \textbf{70}, 3651 (1993).

\bibitem {Stone} M. B. Stone, W. Tian, M. D. Lumsden, G. E. Granroth, D.  Mandrus, J. H. Chung, N. Harrison, S. E. Nagler, Quantum Spin Correlations in an Organometallic Alternating-Sign Chain, Phys. Rev. Lett. \textbf{99}, 087204 (2007).

\bibitem {Bloch12} Immanuel Bloch, Jean Dalibard and Sylvain Nascimb\`{e}ne, Quantum simulations with ultracold quantum gases, Nat. Phys. \textbf{8}, 267 (2012).

\bibitem {Sun14}G. Sun and T. Vekua, Topological order-by-disorder in orbitally degenerate dipolar bosons on a zigzag lattice, Phys. Rev. B \textbf{90}, 094414 (2014).

\bibitem {Sun2014}G. Sun, A. K. Kolezhuk, L. Santos, and T. Vekua, Ferromagnetic spin-orbital liquid of dipolar fermions in zigzag lattices, Phys. Rev. B \textbf{89}, 134420 (2014).

\bibitem {Ohadi17} H. Ohadi, A. J. Ramsay, H. Sigurdsson, Y. delValle-Inclan Redondo, S. I. Tsintzos, Z. Hatzopoulos, T. C. H. Liew, I. A. Shelykh, Y. G. Rubo, P. G. Savvidis, and J. J. Baumberg, Phys. Rev. Lett. \textbf{119}, 067401 (2017).

\bibitem{Daniel} Daniel C. Cabra and Marcelo D. Grynberg, Ground-state magnetization of polymerized spin chains, Phys. Rev. B \textbf{59}, 119 (1999).

\bibitem{Honecker}A. Honecker, Strong-coupling approach to the magnetization process of polymerized quantum spin chains, Phys. Rev. B \textbf{59}, 6790 (1990).

\bibitem{langer} S. Langer, R. Darradi, F. Heidrich-Meisner, W. Brenig, Field-dependent spin and heat conductivities of dimerized spin-1/2 chains, Phys. Rev. B \textbf{82}, 104424 (2010).

\bibitem{Saeed}Saeed Mahdavifar, and Alireza Akbari, Alternating Heisenberg Spin-$1/2$ Chains in a Transverse Magnetic Field, J. Phys. Soc. Jpn. \textbf{77}, 024710 (2008).

\bibitem{Matsuda}M. Matsuda, H. Ueda, A. Kikkawa, Y. Tanaka, K. Katsumata, Y. Narumi, T. Inami, Y. Ueda, and S.-H. Lee, Order-by-disorder and spiral spin-liquid in frustrated diamond-lattice antiferromagnets. Nat. Phys.  \textbf{3}, 397 (2007).

\bibitem{Matsuda2010}     M.  Matsuda,  K.  Ohoyama,  S.  Yoshii,  H.  Nojiri,  P.  Frings,  F. Duc,  B.  Vignolle,  G.  L.  J.  A.  Rikken,  L.-P.  Regnault,  S.-H. Lee,  H.  Ueda,  and  Y.  Ueda, Universal Magnetic Structure of the Half-Magnetization Phase in Cr-Based Spinels, Phys. Rev. Lett. \textbf{104}, 047201 (2010).

\bibitem{You07}  W.-L You, Y.-W. Li, and S.J. Gu, Fidelity, dynamic structure factor, and susceptibility in critical phenomena, Phys. Rev. E \textbf{76}, 022101 (2007).

\bibitem{Gu} S. J. Gu, Fidelity Approach to Quantum Phase Transitions, Int. J. Mod. Phys. B \textbf{24}, 4371 (2010).

\bibitem{You11}W.-L You and Y.-L Dong, Fidelity susceptibility in two-dimensional spin-orbit models, Phys. Rev. B \textbf{84}, 174426  (2011).

\bibitem{You15a}W.-L You, P. Horsch and A, M. Ole\'s, Quantum entanglement in the one-dimensional spin-orbital SU(2)$\otimes XXZ$ model, Phys. Rev. B \textbf{92}, 054423 (2015).

\bibitem{You15b}W.-L You, P. Horsch and A, M. Ole\'s, Entanglement Driven Phase Transitions in Spin-Orbital Models, New J. Phys   \textbf{17}, 083009 (2015).

\bibitem{Chen} Yan Chen, Paolo Zanardi, Z D Wang, and F C Zhang, Sublattice entanglement and quantum phase transitions in antiferromagnetic spin chains, New Journal of Physics \textbf{8}, 97 (2006).

\bibitem{Ren} J. Ren, S. Zhu, Entanglement, fidelity, and quantum phase transition in antiferromagnetic-ferromagnetic alternating Heisenberg chain, Eur Phys J D, \textbf{50}, 103(2008).

\bibitem{Ren02}J. Ren, Y. M. Wang, W. L. You, Quantum phase transitions in spin-1 XXZ chains with rhombic single-ion anisotropy, Phys. Rev. A, \textbf{97}, 0423181(2018).

\bibitem{Venuti10} Lorenzo Campos Venuti and Marco Roncaglia, Equivalence between XY and dimerized models Phys. Rev. A \textbf{81}, 060101R(2010).

\bibitem{Taylor85} J. H. Taylorand G. M\"{u}ller 1985 Physica A 130 1 and references therein.

\bibitem{white} S. R. White, Density-matrix algorithms for quantum renormalization groups, Phys. Rev. B \textbf{48}, 10345 (1993).

\bibitem{U01} U. Schollw\"{o}ck, The density-matrix renormalization group, Rev. Mod. Phys. \textbf{77}, 259 (2005).

\bibitem{vidal01}G. Vidal, Classical Simulation of Infinite-Size Quantum Lattice Systems in One Spatial Dimension, Phys. Rev. Lett. \textbf{98}, 070201 (2007).

\bibitem{vidal02}R Or\'{u}s and G. Vidal, Infinite time-evolving block decimation algorithm beyond unitary evolution, Phys. Rev. B \textbf{78}, 155117 (2008).

\bibitem{zhou}H. Q. Zhou, R. Orus, G. Vidal, Ground state fidelity from tensor network representations, Phys Rev Lett. \textbf{100}, 080601 (2008).

\bibitem{Wang}H. L. Wang, J. H. Zhao, B. Li, H. Q. Zhou, Kosterlitz-Thouless phase transition and ground state fidelity: a novel perspective from matrix product states, Journal of Statistical Mechanics: Theory and Experiment, L10001 (2011).

\bibitem{Zhou2018} Huan-Qiang Zhou, Qian-Qian Shi, and Yan-Wei Dai, Fidelity mechanics: analogues of the four thermodynamic laws and Landauer\'s principle, arXiv:1709.09838.

\bibitem{Wootters}W. K. Wootters, Entanglement of Formation of an Arbitrary State of Two Qubits, Phys. Rev. Lett. \textbf{80} 2245(1998).

\bibitem{Uhrig} G. S. Uhrig, F. Sch\"{o}nfeld, M. Laukamp, and E. Dagotto, Unified quantum mechanical picture for confined spinons
in dimerized and frustrated spin S = 1/2 chains, Eur. Phys. J. B \textbf{7}, 67 (1999).

\bibitem{Chitra} R. Chitra, S. Pati, H. R. Krishnamurthy, D. Sen, and S. Ramasesha, Density-matrix renormalization-group studies of the spin-1/2 Heisenberg system with dimerization and frustration, Phys. Rev. B \textbf{52}, 6581 (1995).

\bibitem{sakai}Toru sakai, Phase Transition of S$=1/2$ Bond-Alternating Chain in a Magnetic Field, J. Phys. Soc. Jpn. \textbf{64}, 251 (1995).

\bibitem{Hida1993} Kazuo Hida, Generalized String Order in the Spin$-1/2$ Alternating Heisenberg Chain, J. Phys. Soc. Jpn. \textbf{62}, 439 (1993).


\bibitem{Lou}Jizhong Lou, Shaojin Qin, and Changfeng Chen, String Order in Half-Integer-Spin Antiferromagnetic Heisenberg Chains, Phys. Rev. Lett. \textbf{91}, 087204 (2003).


\bibitem{Liu01} Jin Hua Liu and Hai Tao Wang, Quantum fidelity, string order parameter, and topological quantum phase transition in a spin-1/2 dimerized and frustrated Heisenberg chain, Eur. Phys. J. B \textbf{8}, 256 (2015).


\bibitem{Ren03}J. Ren, L. P. Gu, W. L. You, Fidelity susceptibility and entanglement entropy in S=1 quantum spin chain with three-site interaction, Acta Physica Sinica, \textbf{67}, 020302 (2018).

\bibitem{Paul}S. Paul, A. K. Ghosh, Ground state properties of the bond alternating spin-1/2 anisotropic Heisenberg chain, Condensed Matter Physics, \textbf{20}, 23701 (2017).

\bibitem{Liu}Guang-Hua Liu, Wen-Long You, Wei Li, and Gang Su, Quantum phase transitions and string orders in the spin-1/2 Heisenberg-Ising alternating chain with Dzyaloshinskii-Moriya interaction, J. Phys.: Condens. Matter \textbf{27}, 165602 (2015).

\bibitem{Calabrese} P. Calabrese and J. Cardy, Entanglement entropy and quantum field theory, J. Stat. Mech (2004) P06002.

\bibitem{Nicolas} N. Laflorencie, E. S. S{\o}rensen, M. S. Chang, and I. Affleck, Boundary Effects in the Critical Scaling of Entanglement Entropy in 1D Systems, Phys. Rev. Lett. \textbf {96}, 100603(2006).

\bibitem{Chiara} G. D. Chiara, S. Montangero, P. Calabrese, and R. Fazio, Entanglement Entropy dynamics in Heisenberg chains, J. Stat. Mech (2006) P03001 [cond-mat/0512586].

\bibitem{Ren01} J. Ren, S. Q. Zhu, X. Hao, Entanglement entropy in an antiferromagnetic Heisenberg spin chain with boundary impurities, J Phys B-at Mol Opt, \textbf{42}, 015504 (2009).

\bibitem{Plenio}T. Baumgratz, M. Cramer, and M. B. Plenio, Quantifying Coherence, Phys. Rev. Lett. \textbf{113}, 140401 (2014).

\bibitem{Zhang} D. J. Zhang, C. L. Liu, X.-D. Yu, D. M. Tong, Estimating Coherence Measures from Limited Experimental Data Available, Phys. Rev. Lett. \textbf{120}, 170501 (2018).


\bibitem{Zheng}W. Zheng, Z. Ma, H. Wang, S.-M. Fei, X. Peng, Experimental Demonstration of Observability and Operability of Robustness of Coherence, Phys. Rev. Lett. \textbf{120}, 230504 (2018).


\bibitem{Girolami} D. Girolami, Observable Measure of Quantum Coherence in Finite Dimensional Systems, Phys. Rev. Lett. \textbf{113}, 170401 (2014).


\bibitem{Karpat}G. Karpat, B. \c{C}akmak, F. F. Fanchini, Quantum coherence and uncertainty in the anisotropic XY chain, Phys. Rev. B \textbf{90}, 104431 (2014).

\end{thebibliography}
\end{document}